\documentclass[a4paper,11pt]{article}
\usepackage{jcappub} % for details on the use of the package, please see the JINST-author-manual
\usepackage{lineno}
\usepackage{graphicx}   % need for figures
\usepackage{verbatim}   % useful for program listings
\usepackage[normalem]{ulem} % strike out without modifying emph
\usepackage{subfigure}  % use for side-by-side figures
\usepackage{hhline}
\usepackage{url}
\usepackage{hyperref}   % use for hypertext links, including those to external documents and URLs
\usepackage{amssymb}
\usepackage{enumerate}	% can be used to enumerate several items
\usepackage{color}

\def\apjl{APJL}
\def\mnras{MNRAS}
\def\apj{APJ}

\title{\boldmath Galaxy Group Spin Alignment with Cosmic Filament in the TNG Simulation}

\begin{document}
% \begin{CJK*}{UTF8}{gbsn}

\author[a,b,c]{Wei Wang*}
\author[c]{Peng Wang*}
\author[b,d]{Yu Rong*}
\author[c,e]{Hao-da Wang}
\author[c,e]{Xiao-xiao Tang}

\affiliation[a]{Purple Mountain Observatory, Chinese Academy of Sciences, No.10 Yuan Hua Road, 210034 Nanjing, China.}
\affiliation[b]{School of Astronomy and Space Science, University of Science and Technology of China, Hefei 230026, Anhui, China.}
\affiliation[c]{Shanghai Astronomical Observatory, Chinese Academy of Sciences, Nandan Road 80, Shanghai 200030, China}
\affiliation[d]{Department of Astronomy, University of Science and Technology of China, Hefei, Anhui 230026, China}
\affiliation[e]{University of Chinese Academy of Sciences, Beijing 100049, China}

% E-mail addresses: only for the corresponding author
% \emailAdd{first@one.univ}
\emailAdd{wangwei@pmo.ac.cn,pwang@shao.ac.cn, rongyua@ustc.edu.cn}

\abstract{We investigate the alignment between the spin vectors of galaxy groups and the axes of their nearest cosmic filaments using the TNG300-1 cosmological hydrodynamical simulation. By systematically analyzing a large sample of groups, we find a robust perpendicular alignment between group spin and filament orientation. Among all examined properties, only group mass and the distance to the nearest filament significantly affect the strength of this alignment: more massive groups and those closer to filaments exhibit a stronger perpendicular signal. In contrast, the alignment is largely insensitive to group richness, the stellar mass threshold used to select member galaxies, and redshift. We further quantify the bias introduced by using member galaxies as tracers of group spin, finding a typical misalignment angle of $\sim38^\circ$ between the spin measured from all dark matter particles and that inferred from member galaxies, independent of group richness or stellar mass cut. Our results provide a clear theoretical benchmark for interpreting observational measurements of spin-filament alignment and highlight the importance of considering group mass and environment. These findings help clarify the main factors influencing spin-filament alignment and provide useful context for future observational and theoretical studies of angular momentum in the cosmic web.}

\maketitle
\flushbottom

\section{Introduction}
\label{sec:intro}

In the standard $\Lambda$CDM cosmological model, the large-scale structure of the Universe emerges from the gravitational amplification of primordial density perturbations\citep{1996Natur.380..603B}, resulting in a cosmic web composed of voids, sheets, filaments, and nodes \citep[e.g.,][]{2007MNRAS.375..489H}. At the intersections of filaments, galaxy groups and clusters-virialized systems containing dozens to thousands of galaxies—form through anisotropic accretion and mergers \citep[e.g.,][]{2005Natur.435..629S,2014MNRAS.444.1518V}. Understanding their formation and dynamical evolution is key to unraveling the history of cosmic structure growth.

A fundamental dynamical quantity of galaxy groups is their angular momentum (or ``spin''), which encapsulates the cumulative effects of past interactions, mergers, and infall of surrounding matter \citep[e.g.,][]{2014MNRAS.443.1274L,2015ApJ...807...37S,2015ApJ...813....6K,2025arXiv250420133A}. The spin vector characterizes the coherent rotation of the system and carries valuable information about its assembly history and the anisotropic environment in which it formed. The orientation and magnitude of group spin influence various properties of the member galaxies, including morphology, star formation activity, and satellite distribution \citep[e.g.,][]{2014MNRAS.445L..46W,2015MNRAS.452.3369C,2014MNRAS.444.1453D}. Thus, the angular momentum of galaxy groups serves not only as a dynamical tracer, but also as a bridge linking the large-scale environment to internal galactic processes.

From a theoretical perspective, dark matter halos acquire angular momentum through gravitational tidal torques exerted by surrounding overdensities during the linear regime of structure formation-a process formalized in the tidal torque theory \citep{Hoyle1949,1969ApJ...155..393P,1970Ap......6..320D,1979MNRAS.186..133E,1984ApJ...286...38W,1987ApJ...319..575B,2007JRSSC..56....1S,2002MNRAS.332..325P,2002MNRAS.332..339P,2021MNRAS.502.5528L,2024PASP..136c7001L,2025arXiv250501298L}. In the subsequent nonlinear stage, angular momentum can also be generated via mergers and anisotropic accretion, particularly along filaments \citep[e.g.,][]{2014MNRAS.443.1274L,2015ApJ...807...37S,2015ApJ...813....6K,2017MNRAS.468L.123W,2018MNRAS.473.1562W}. Cosmological simulations have revealed a characteristic mass-dependent behavior of spin-filament alignments: low-mass halos tend to have spins aligned parallel to filament spines, while high-mass halos more frequently exhibit perpendicular alignments \citep[e.g.,][]{2012MNRAS.427.3320C,2014MNRAS.444.1453D,2018MNRAS.473.1562W,2018MNRAS.481..414G,2021MNRAS.503.2280G,2021NatAs...5..742K}. This spin-flip transition reflects the changing balance between smooth accretion and merger-driven spin acquisition as structures evolve.

Observationally, studies using SDSS data have begun to test these predictions. For example, the spin axes of spiral galaxies and dwarf galaxies have been found to preferentially align parallel to nearby filament spines \citep{2013ApJ...775L..42T,2015MNRAS.450.2727T,2015ApJ...798...17Z,2020MNRAS.498L..72R}, consistent with expectations for low-mass halos. Based on recent observational works from  integral field unit (IFU) spectroscopy sky survey, alignment signal still can be found for low stellar mass spiral galaxies and lower bulge mass galaxies in MaNGA\citep{2019ApJ...876...52K,2021MNRAS.504.4626K,2025ApJ...987L..30W} and SAMI\citep{2020MNRAS.491.2864W,2022MNRAS.516.3569B,2023MNRAS.526.1613B} galaxy survey. For galaxy groups, however, the alignment trend differs. Recently, \cite{2025ApJ...983L...3R} analyzed over 30,000 galaxy groups from SDSS DR12 and found a statistically significant perpendicular alignment between the projected angular momentum vectors of galaxy groups and the projected directions of their host filament spines. This result supports a scenario in which the spin of galaxy groups arises from the orbital angular momentum of infalling member galaxies, which tend to accrete along filament directions.

Nevertheless, observational approaches are fundamentally limited by projection effects, redshift-space distortions, and uncertainties in group membership and velocity estimates. In particular, the ``Fingers of God'' effect complicates the interpretation of redshift-based distances and motions, potentially biasing the inferred spin-filament alignment signals. Moreover, observational spin estimates often rely on simplified assumptions such as weighting by stellar mass or galaxy distance from the group center.

Cosmological hydrodynamical simulations offer a crucial complement to observations by providing full 3D information about mass distributions and velocity fields. The IllustrisTNG project \citep{2018MNRAS.480.5113M,2018MNRAS.477.1206N,2018MNRAS.475..624N,2018MNRAS.475..648P,2018MNRAS.475..676S} combines large volumes with high resolution, enabling detailed studies of the angular momentum of groups and their relation to the cosmic web in a controlled $\Lambda$CDM environment.

In this study, we use the TNG300-1 simulation from the IllustrisTNG \citep{2018MNRAS.480.5113M,2018MNRAS.477.1206N,2018MNRAS.475..624N,2018MNRAS.475..648P,2018MNRAS.475..676S} suite to revisit and extend the findings of \citep{2025ApJ...983L...3R}. Specifically, we aim to: (1) reconstruct the 3D spin vectors of galaxy groups and quantify their alignment with nearby filament spines; (2) investigate how the alignment signal depends on group mass, richness, and internal structure; and (3) assess to what extent the group spin estimated from member galaxies can reliably trace the true spin of the underlying dark matter halo, given that direct measurements of group spin are not feasible in observations and member galaxies are commonly used as tracers. By connecting observations with simulations, this work provides new insight into the reliability and limitations of using member galaxies as tracers of group spin, and clarifies the implications for interpreting observed spin-filament alignments in the context of large-scale structure.

The structure of this paper is as follows. In Section~2, we describe the simulation data and methods for measuring group spin and filament orientation. Section~3 presents the alignment results and their dependence on group properties. In Section~4 and Section~5, we summarize and discuss our main conclusions.

%%%%%%%%%%%%%%%%%%%%%%%%%%%%%%%%%%%%%%%%%%%%%%%%%%%%%
%%%%%%%%%%%%% 
%Figure 1 
%%%%%%%%%%%%%

%%%%%%%%%%
%   Method
%%%%%%%%%%
\section{Data and Methodology}
\label{sec:method}

\subsection{Simulation Overview and sample selection}

Our analysis is based on the TNG300-1 simulation, the largest volume of the IllustrisTNG project \citep{2018MNRAS.480.5113M,2018MNRAS.477.1206N,2018MNRAS.475..624N,2018MNRAS.475..648P,2018MNRAS.475..676S}, which follows the evolution of dark matter, baryons, and black holes in a $(300~\mathrm{Mpc})^3$ cosmological volume using the moving-mesh code \texttt{AREPO} \citep{2010MNRAS.401..791S}. The simulation adopts a $\Lambda$CDM cosmology consistent with \textit{Planck 2015} parameters \citep{2016A&A...594A..13P}: $\Omega_\mathrm{m} = 0.3089$, $\Omega_\Lambda = 0.6911$, $\Omega_\mathrm{b} = 0.0486$, $H_0 = 67.74~\mathrm{km~s^{-1}~Mpc^{-1}}$, and $\sigma_8 = 0.8159$. The dark matter and baryonic mass resolutions are $m_\mathrm{DM} \approx 5.9 \times 10^7~M_\odot$ and $m_\mathrm{b} \approx 1.1 \times 10^7~M_\odot$, respectively.

Dark matter halos (Groups) are identified using the standard Friends-of-Friends (FoF) algorithm \citep{1985ApJ...292..371D}, while the SUBFIND algorithm \citep{2001MNRAS.328..726S,2009MNRAS.399..497D} is applied to each FoF group to identify gravitationally bound structures. The group mass, $\rm M_{200}$, used here is the total mass of this group enclosed in a sphere with mean density being 200 times the critical density of the Universe.

\begin{figure}[!htp]
\centering
\includegraphics[width=0.96\linewidth]{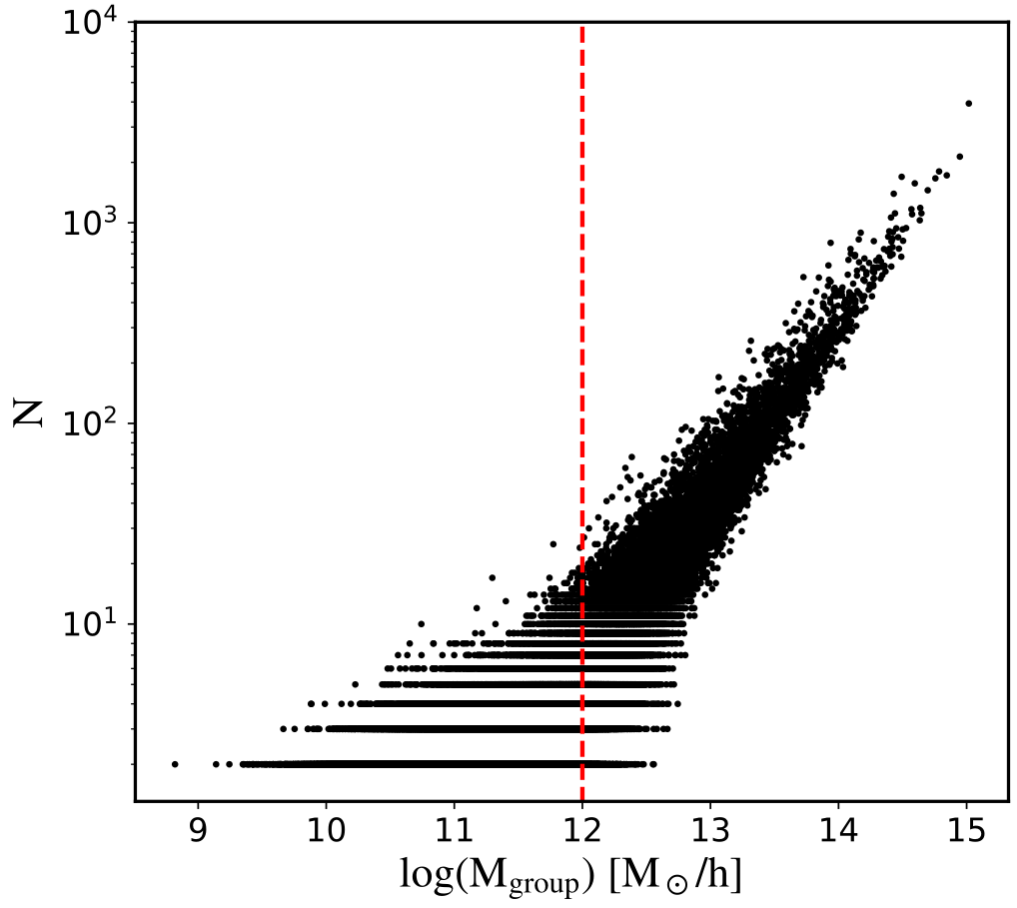}
\caption{
The number of member galaxies ($N$) as a function of the logarithm of group mass, $\rm \log(M_\mathrm{group})\ [M_\odot/h]$. The plot illustrates how the number of member galaxies increases with group mass, where group mass is given in solar masses on a logarithmic scale. The red short line indicates the lower mass limit for groups used in our analysis, set at $\rm 10^{12}\ M_\odot/h$.
}
\label{fig:f1}
\end{figure}

Based on the general understanding of group mass and the relationship between group mass and member galaxy number illustrated in Figure~\ref{fig:f1}, we set \(\rm 10^{12}\ M_\odot/h\) as the lower mass limit for group selection. Figure~\ref{fig:f1} shows that the number of member galaxies (\(N\), richness) increases with group mass, and this correlation becomes notably tighter above \(M_\mathrm{group} = 10^{12}\ M_\odot\), as indicated by the red short line. Below this threshold, numerical effects from simulation resolution are more significant. This selection ensures that only well-defined, massive groups are included in our subsequent alignment analysis.

%%%%%%%%%%%%% 
%   Fig. 1
%%%%%%%%%%%%%

\subsection{Filament Identification}

Filaments are identified using the Discrete Persistent Structures Extractor \citep[hereafter \texttt{DisPerSE;}][]{2011MNRAS.414..350S,2011MNRAS.414..384S} based on the galaxy distribution by adopting the parameter settings and methodology of \citep{2024MNRAS.532.4604W}, selecting galaxies with stellar mass \( M_* > 10^9\, M_\odot \) to trace filamentary structures. As discussed in \citep{2024MNRAS.532.4604W}, the choice of stellar mass threshold has little affect on the boundary of filaments because of vanishing of weak filaments in high mass cut. In this paper, we adopt the choice of mass threshold following the observational limits of \citep{2004MNRAS.351.1151B,2011MNRAS.418.1587T}.

After tracing the filament, we calculate the distance from groups to their nearest filament segment. The orientation of each filament is given by the local tangent vector of the spine, denoted $\mathbf{e}_\mathrm{f}$, at the closest point to the group center.

\subsection{Group Angular Momentum, Alignment Angle and Statistical Analysis}

The 3D angular momentum of each galaxy group is computed using its member galaxies. The spin vector is defined as 
\begin{equation}
    \mathbf{L} = \sum_{i=1}^{N} m_{*,i} (\mathbf{r}_i - \mathbf{r}_\mathrm{c}) \times (\mathbf{v}_i - \mathbf{v}_\mathrm{c}),
\end{equation}
where $m_{*,i}$, $\mathbf{r}_i$, and $\mathbf{v}_i$ are the stellar mass, position, and velocity of the $i$-th galaxy, and $\mathbf{r}_\mathrm{c}$ and $\mathbf{v}_\mathrm{c}$ refer to the position and velocity of the central (most massive) galaxy in the group. To facilitate direct comparison with observations, we adopt the stellar mass of each galaxy in our fiducial calculation. We have also tested using the total subhalo mass, and find that this choice does not significantly affect our results.

The alignment between the group angular momentum $\mathbf{L}$ and filament direction $\mathbf{e}_\mathrm{f}$ is quantified via the angle:
\begin{equation}
    \cos\theta = \frac{\mathbf{L} \cdot \mathbf{e}_\mathrm{f}}{|\mathbf{L}||\mathbf{e}_\mathrm{f}|},
\end{equation}
The alignment angles are restricted in the range of $[0^\circ, 90^\circ]$,
where $\theta \sim 90^\circ$ indicates perpendicular alignment and $\theta \sim 0^\circ$ indicates parallel alignment.

To evaluate the significance of the alignment signal, specifically the extent of its deviation from randomness, we performed the following calculations and tests. The calculations are carried out starting from a theoretical derivation. Let two unit vectors be independently and uniformly distributed on the unit sphere \( S^2 \). The probability density function (PDF) of the angle \( \theta \in [0, \pi] \) between them is given by
\[
p(\theta) = \frac{1}{2} \sin \theta.
\]
This arises from the fact that the differential area element on the sphere is \( \sin \theta \, d\theta \, d\phi \), and integrating out the azimuthal angle yields a distribution proportional to \( \sin \theta \). As mentioned above, we restrict our attention to pairs of vectors where the angle \( \theta \in [0, \frac{\pi}{2}] \). In this interval, the original PDF must be renormalized. The normalization factor is
\[
\int_0^{\frac{\pi}{2}} \frac{1}{2} \sin \theta \, d\theta = \frac{1}{2} [-\cos \theta]_0^{\frac{\pi}{2}} = \frac{1}{2} (1 - 0) = \frac{1}{2},
\]
so the normalized PDF becomes
\[
p^*(\theta) = \frac{1}{2} \sin \theta \bigg/ \frac{1}{2} = \sin \theta.
\]
The expected angle over \( [0, \frac{\pi}{2}] \) is then
\[
\mathbb{E}[\theta] = \int_0^{\frac{\pi}{2}} \theta \cdot \sin \theta \, d\theta.
\]
To evaluate this, we use integration by parts. Let \( u = \theta \), \( dv = \sin \theta \, d\theta \), then \( du = d\theta \), \( v = -\cos \theta \). This gives
\[
\int \theta \sin \theta \, d\theta = -\theta \cos \theta + \int \cos \theta \, d\theta = -\theta \cos \theta + \sin \theta.
\]
Evaluating the definite integral:
\begin{align}
\mathbb{E}[\theta] & = \left[ -\theta \cos \theta + \sin \theta \right]_0^{\frac{\pi}{2}} \nonumber \\ 
& = -\frac{\pi}{2} \cdot \cos\left(\frac{\pi}{2}\right) + \sin\left(\frac{\pi}{2}\right) - (0 + \sin 0) \nonumber \\
%& = 0 + 1 \nonumber \\
& = 1 \text{ rad} \approx 57.3^\circ.
\label{equ:rand}
\end{align}
Therefore, the expected angle between two random unit vectors in 3D, conditioned on their angle lying in \( [0, \frac{\pi}{2}] \) is $\approx 57.3^\circ$. If a measured angle is larger than $57.3^\circ$, we refer it to as a perpendicular trend.  If a measured angle smaller than $57.3^\circ$, we refer it to as a align trend.

For the random tests, we generate 100 random samples. Each random sample contains the same number of groups as the TNG sample. In each random trial, the spin vector of each group is randomly assigned while their hosted filament direction is kept fixed. Then, we calculate the mean value of $\theta_R$ for each random sample, resulting in a distribution of 100 mean values of $\theta_R$. The standard deviation of this distribution is then used to quantify the significance of the alignment signal in TNG via 
\[
\mathrm{significance} = \frac{|\theta_{\rm TNG} - \theta_R|}{\sigma(\theta_R)}
\]

In addition, to quantify alignment strength, we also follow the work of \cite{2025ApJ...983L...3R} and \cite{2025ApJ...983L...3R} to define an alignment index:
\[
\mathcal{I} = \frac{N_{0^\circ\text{--}57.3^\circ}}{N_{57.3^\circ\text{--}90^\circ}}
\]
where $N_{x^\circ\text{--}y^\circ}$ is the number of groups with alignment angle $\theta$ in the corresponding range. The smaller the value of the $\mathcal{I}$, the stronger the perpendicular alignment signal. The uncertainty on the $\mathcal{I}$ is obtained from the random trials.

%%%%%%%%%%%%%%%%%%%%%%%%%%%%%%%%%%%%%%%%%%%%%%%%%%%%%
%%%%%%%%%%%%% 
%   Result
%%%%%%%%%%%%% 
\section{Results}
\label{sec:result}

%%%%%%%%%%%%% 
%   Fig. 2
%%%%%%%%%%%%% 
\begin{figure}[!htp]
\includegraphics[width=0.96\linewidth]{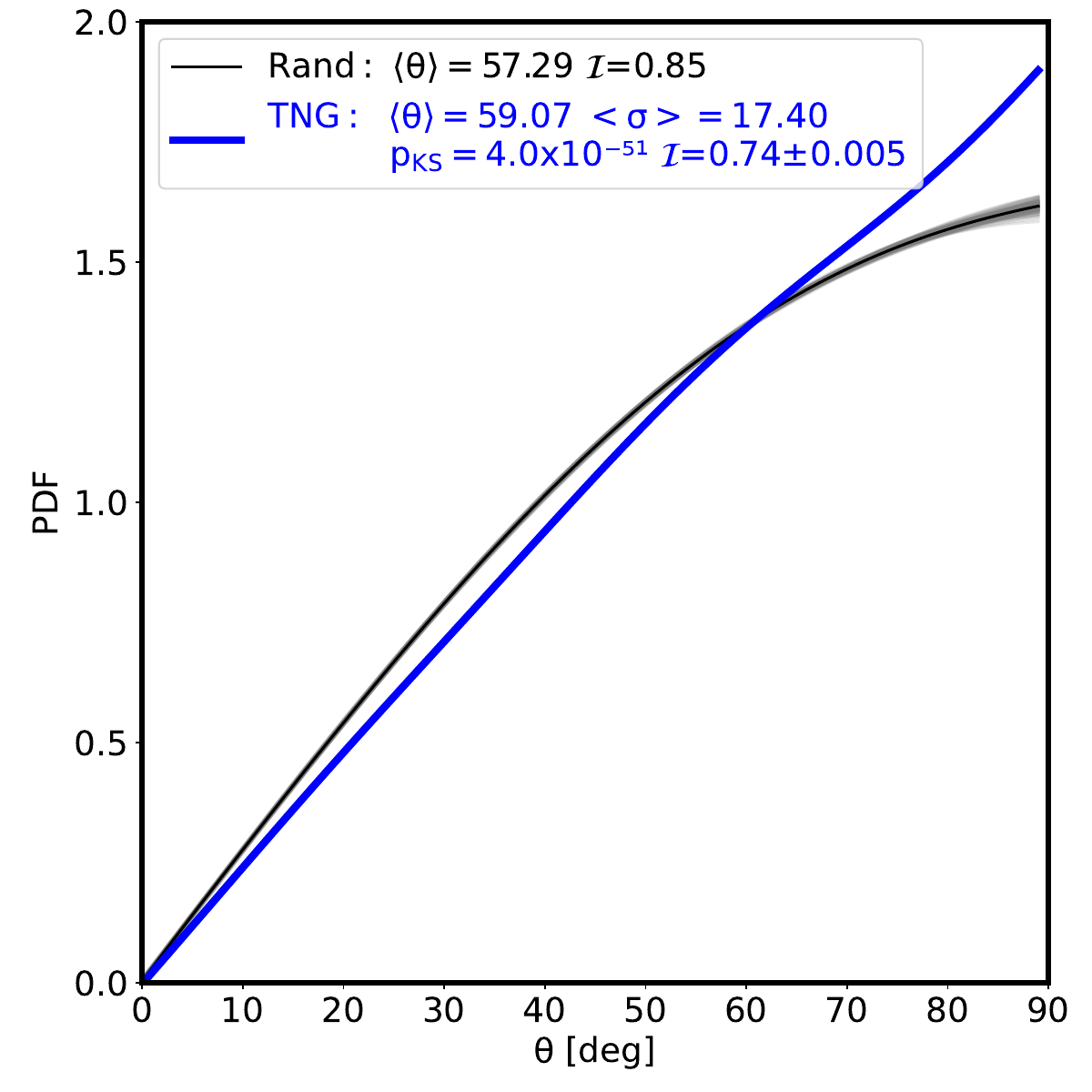}
\caption{The probability density function (PDF) of the angle $\theta$ between the spin vectors of galaxy groups and the axes of their nearest cosmic filaments, with $\theta \in [0^\circ, 90^\circ]$, is shown. The TNG sample, represented by the blue solid line with $\langle\theta\rangle = 59.07^\circ$, is compared against 100 randomized trials with $\langle\theta\rangle = 57.29^\circ$ (consistent with the conclusion of Eq.~\ref{equ:rand}). The mean of the random trials is shown as a black solid line. The perpendicular alignment signal exhibits a significance of $17.40\sigma$, indicating a strong deviation of the TNG measurement from the null hypothesis of random orientation. A Kolmogorov-Smirnov test yields a probability of $p_{\mathrm{KS}} = 4.0 \times 10^{-51}$, confirming a highly significant difference between the two distributions. This strong perpendicular alignment can also be seen from the alignment index, $\mathcal{I} = 0.74 \pm 0.005$, which is less than the alignment index($\mathcal{I}=0.85$) in random trials.}
\label{fig:f2}
\end{figure}

In this section, we present the main results of our analysis on the alignment between the spin vectors of galaxy groups and the axes of their nearest cosmic filaments in the TNG300-1 simulation.

%%%%%%%%%%%%% 
%   Fig. 3
%%%%%%%%%%%%% 
\begin{figure*}[!htp]
% \plotthree{f3a.pdf}{f3b.pdf}{f3c.pdf}
\subfigure{\includegraphics[width=0.31\linewidth]{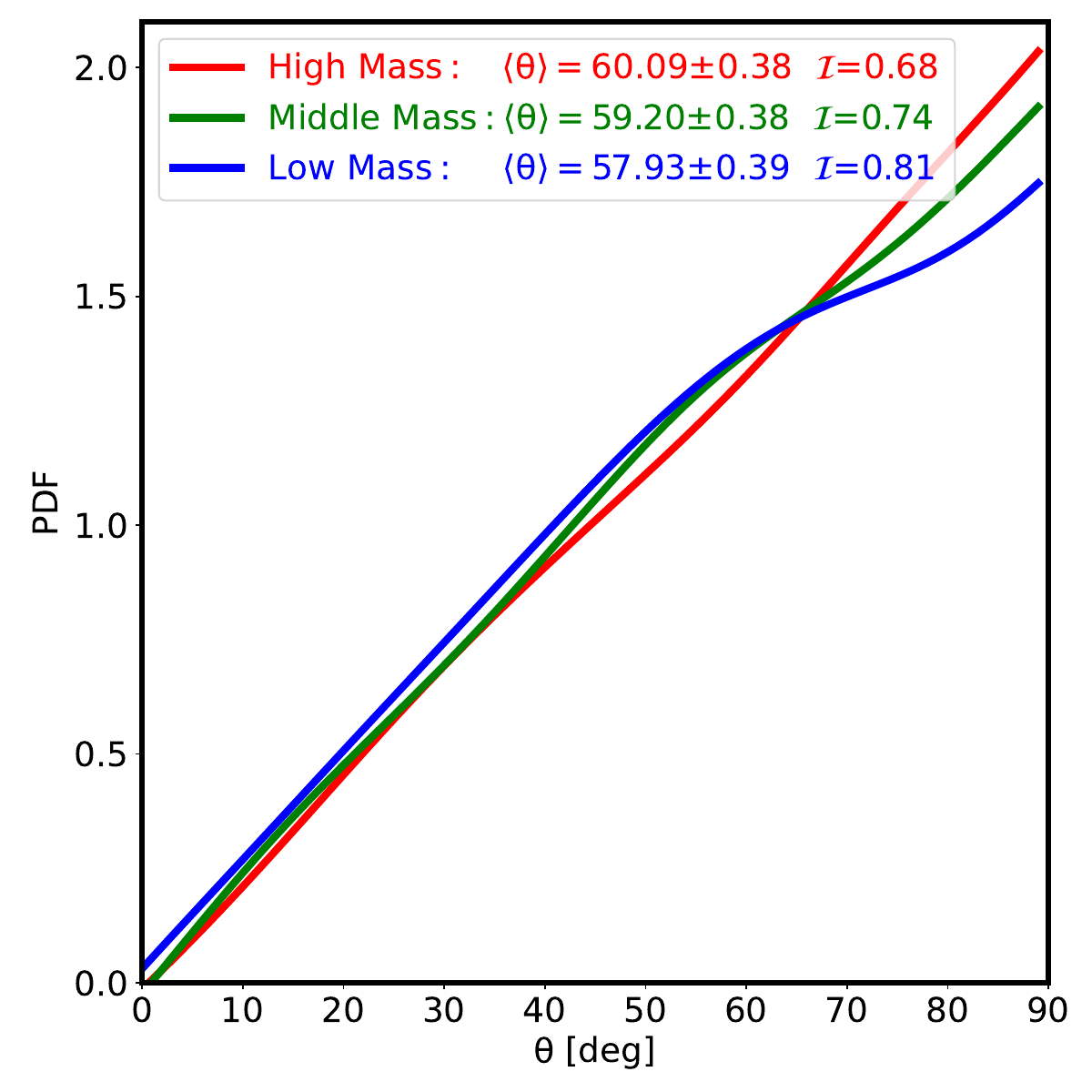}}
\subfigure{\includegraphics[width=0.31\linewidth]{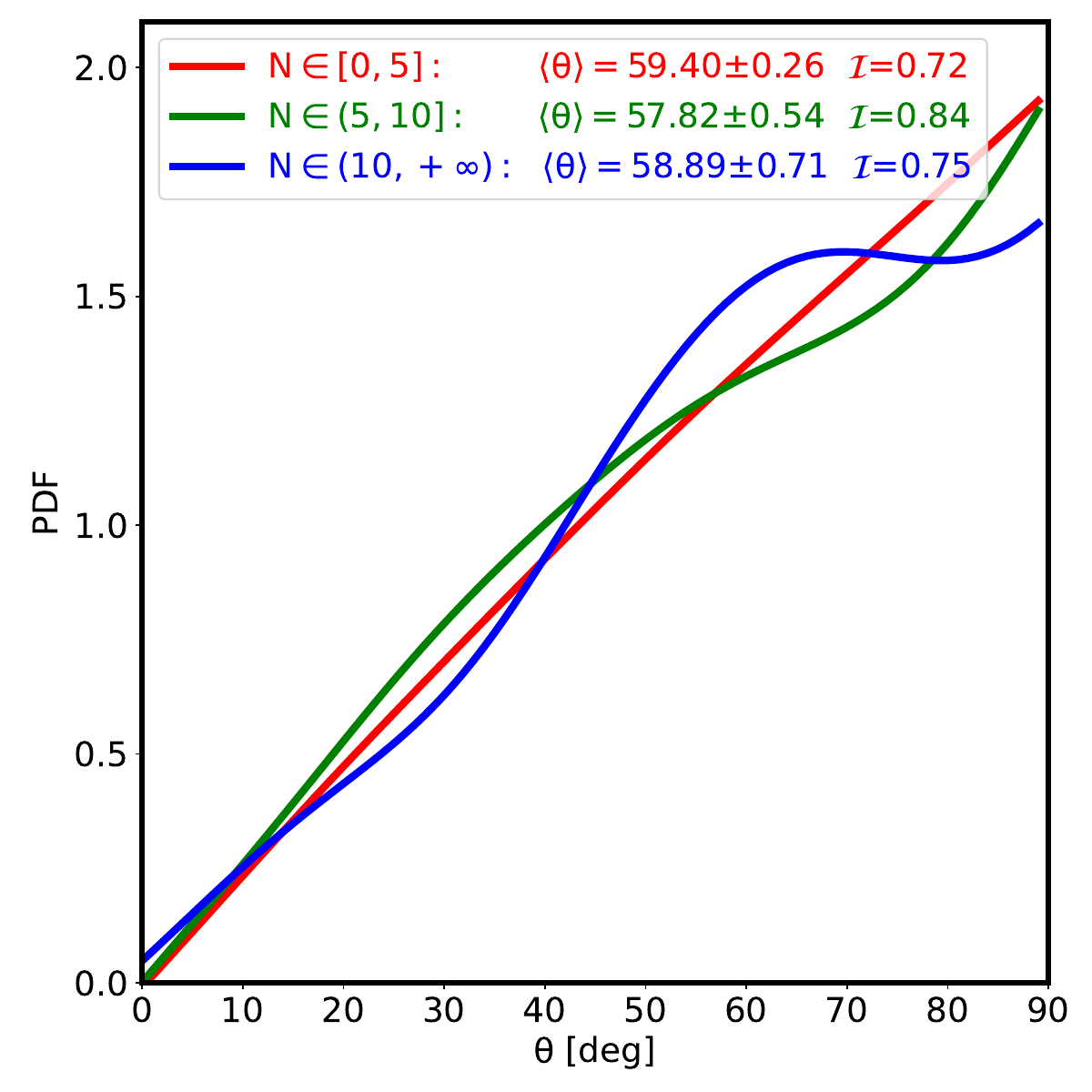}}
\subfigure{\includegraphics[width=0.31\linewidth]{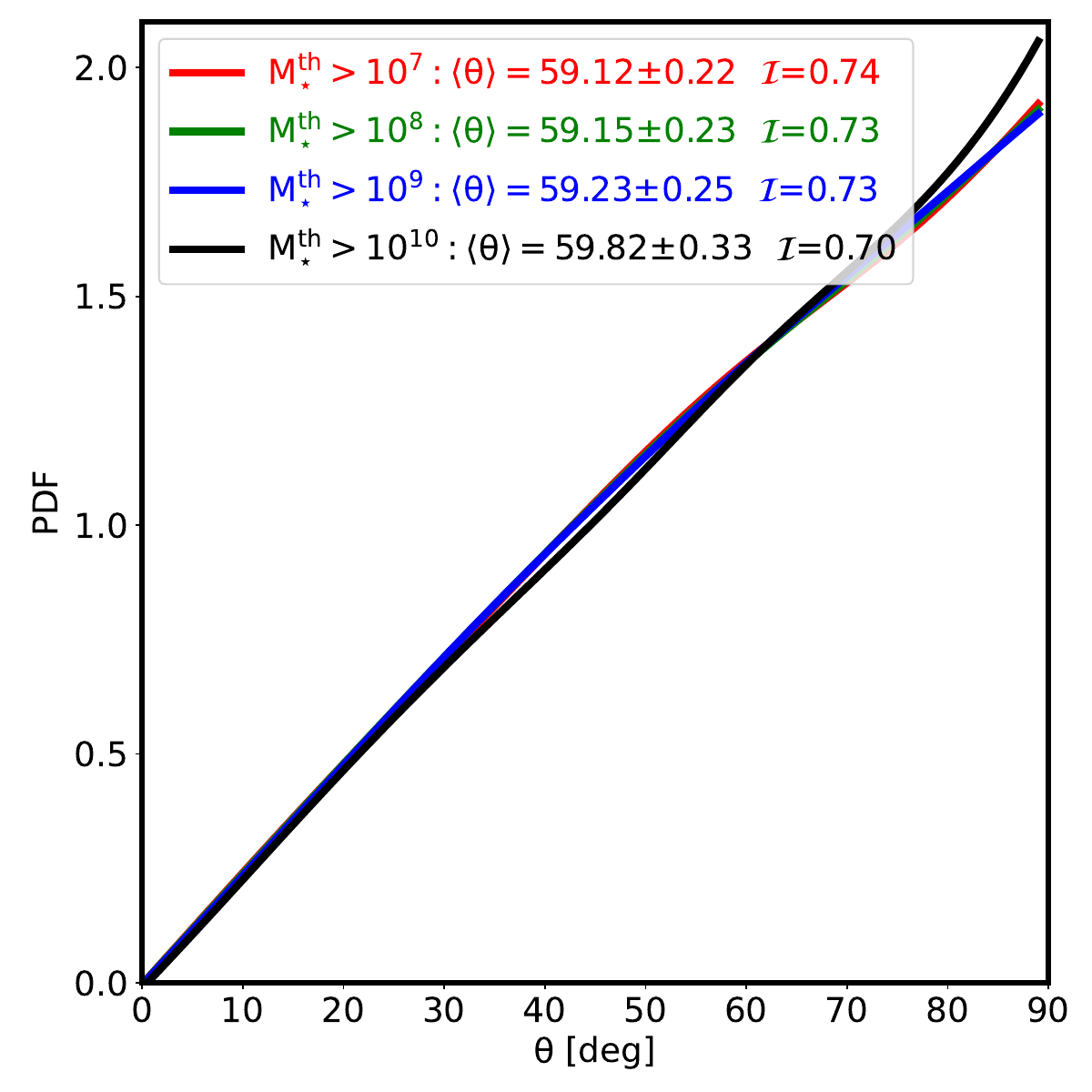}}
\caption{
The probability density functions (PDFs) of the alignment angle $\theta$ between the spin vectors of galaxy groups and the axes of their nearest filaments, shown for different subsamples. \textbf{Left:} Groups divided by mass (high, middle, and low mass bins) with same sample size. \textbf{Middle:} Groups divided by the number of member galaxies ($N$). \textbf{Right:} Groups with different stellar mass thresholds ($M_*^{\rm th}$) for member galaxies when estimating group spin. Different subsamples are shown in different colors; the mean alignment angle $\langle\theta\rangle$ and alignment index $\mathcal{I}$ for each are given in the legend, respectively.
}
\label{fig:f3}
\end{figure*}

\subsection{Overall Perpendicular Alignment Signal}
Building upon the methodology described in Section~\ref{sec:method}, we first examine the overall distribution of alignment angles and compare it to the expectation from random samples.

Figure~\ref{fig:f2} shows the probability density function (PDF) of the angle $\theta$ between the spin vectors of galaxy groups and the directions of their nearest filaments, restricted to the range $\theta \in [0^\circ, 90^\circ]$. The blue solid line represents the TNG sample, while the black solid line corresponds to the mean result from 100 randomized trials, which serve as the null hypothesis of random orientation. The mean alignment angle for the TNG groups is $\langle\theta\rangle = 59.07^\circ$, which is significantly larger than the random expectation of $\langle\theta\rangle = 57.29^\circ$ (also see Eq.~\ref{equ:rand}). This indicates a clear tendency for the spin vectors of galaxy groups to be preferentially perpendicular to the axes of their host filaments.

The statistical significance of this deviation is remarkable: the perpendicular alignment signal in the TNG sample differs from the random case at the $17.40\sigma$ level. Furthermore, a Kolmogorov-Smirnov test yields a probability of $p_{\mathrm{KS}} = 4.0 \times 10^{-51}$, confirming that the observed distribution is highly inconsistent with the null hypothesis of random alignment. This strong alignment is also reflected in the alignment index, $\mathcal{I} = 074 \pm 0.005$, which quantifies the excess of perpendicular alignments.   
These results provide robust evidence that, in the TNG300-1 simulation, the angular momentum of galaxy groups is not randomly oriented with respect to the cosmic web, but instead exhibits a strong perpendicular alignment with nearby filamentary structures.

\subsection{Dependence of Alignment on Group Properties}

To further investigate the robustness of the perpendicular alignment signal seen in Figure~\ref{fig:f2}, we examine how the alignment depends on group mass, richness, and the choice of stellar mass threshold for member galaxies. Figure~\ref{fig:f3} presents the probability density functions (PDFs) of the alignment angle $\theta$ for various subsamples. For each subsample, the mean alignment angle $\langle\theta\rangle$ and alignment index $\mathcal{I}$ are indicated in the legend.

In the left panel, groups are divided into three mass bins, each containing an equal number of groups. It is evident that the perpendicular alignment becomes more pronounced with increasing group mass. The mean alignment angles for the high-, middle-, and low-mass bins are $\langle \theta \rangle = 60.09^\circ \pm 0.38$, $59.20^\circ \pm 0.38$, and $57.93^\circ \pm 0.39$, with corresponding alignment indices of $\mathcal{I} = 0.68$, $0.74$, and $0.81$, respectively. The highest mass bin therefore exhibits the largest mean angle and the smallest alignment index, indicating a stronger excess of perpendicular alignments relative to the random expectation. This trend suggests that more massive groups, which are more likely to have experienced frequent mergers and anisotropic accretion along filaments, develop a more perpendicular spin orientation relative to the surrounding large-scale structure. This scenario is consistent with previous studies \citep[e.g.,][]{2014MNRAS.443.1274L,2015ApJ...807...37S,2015ApJ...813....6K,2018MNRAS.473.1562W}.

The middle panel shows the alignment PDFs for groups binned by the number of member galaxies ($N$). Groups are divided into three bins with $N \in [0,5]$, $N \in (5,10]$, and $N \in (10,+\infty)$. The corresponding mean alignment angles are $\langle \theta \rangle = 59.40^\circ \pm 0.26$, $57.82^\circ \pm 0.54$, and $58.89^\circ \pm 0.71$, with alignment strengths of $\mathcal{I} = 0.72$, $0.84$, and $0.75$, respectively. These values fluctuate slightly across the bins but do not reveal any significant or monotonic trend.

The right panel shows the alignment PDFs for groups with different stellar mass thresholds applied to member galaxies when estimating group spin. Specifically, for four thresholds of $M^{\rm th} > 10^7$, $10^8$, $10^9$, and $10^{10}\,M_\odot$, the mean alignment angles are $\langle \theta \rangle = 59.12^\circ \pm 0.22$, $59.15^\circ \pm 0.23$, $59.23^\circ \pm 0.25$, and $59.82^\circ \pm 0.33$, with corresponding alignment indices of $\mathcal{I} = 0.74$, $0.73$, $0.73$, and $0.70$, respectively. These results demonstrate that the perpendicular alignment signal is robust against the choice of stellar mass threshold, with both $\langle\theta\rangle$ and $\mathcal{I}$ remaining nearly constant across the different cuts.

%%%%%%%%%%%%% 
%   Fig. 4
%%%%%%%%%%%%% 
\begin{figure*}[!htp]
% \plottwo{f4a.pdf}{f4b.pdf}
\subfigure{\includegraphics[width=0.46\linewidth]{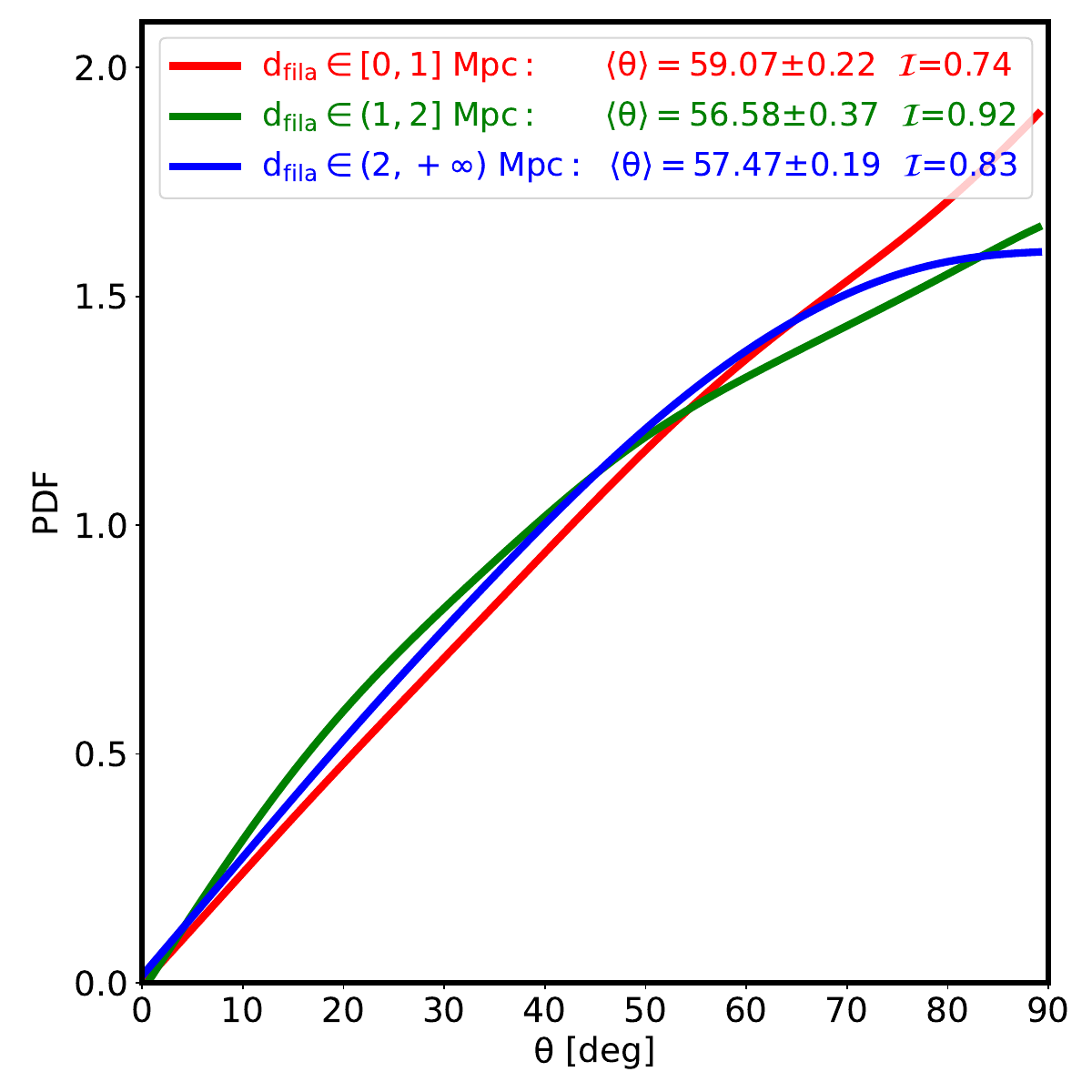}}
\subfigure{\includegraphics[width=0.46\linewidth]{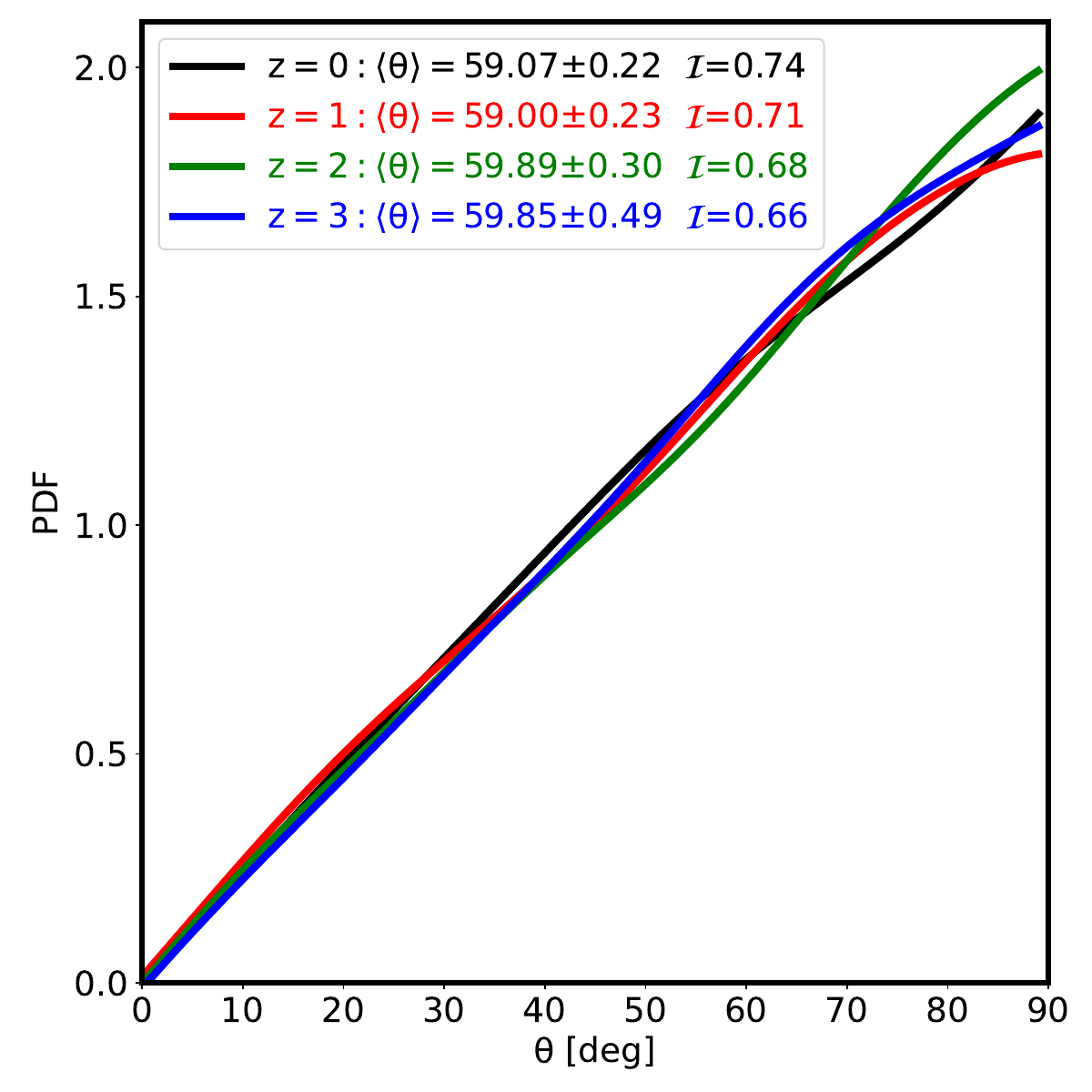}}
\caption{\textbf{Left:} The probability density functions (PDFs) of the alignment angle $\theta$ between the spin vectors of galaxy groups and the axes of their nearest filaments, for different ranges of group-filament distance $d_\mathrm{fila}$. The red, green, and blue curves correspond to groups with $d_\mathrm{fila} \in [0, 1]$~Mpc, $d_\mathrm{fila} \in (1, 2]$~Mpc, and $d_\mathrm{fila} \in (2, +\infty)$~Mpc, respectively. For each subsample, the mean alignment angle $\langle\theta\rangle$ and alignment index $\mathcal{I}$ are indicated in the legend.
\textbf{Right:} Groups at different redshifts ($z = 0, 1, 2, 3$). For each subsample, the mean alignment angle $\langle\theta\rangle$ and alignment index $\mathcal{I}$ are indicated in the legend. In all cases, a preference for perpendicular alignment is observed, with the strength of the signal varying systematically with group mass, richness, and redshift.
}
\label{fig:f4}
\end{figure*}

We further explore how the alignment signal depends on the distance between galaxy groups and their nearest filaments, as well as its evolution with cosmic time. 
Figure~\ref{fig:f4} shows the probability density functions (PDFs) of the alignment angle $\theta$ for different subsamples.

The left panel presents the alignment PDFs for groups in three bins of the distance between group and filament $d_\mathrm{fila}$. 
The red, green, and blue curves correspond to groups with 
$d_\mathrm{fila} \in [0,1]$~Mpc, $d_\mathrm{fila} \in (1,2]$~Mpc, and $d_\mathrm{fila} \in (2,+\infty)$~Mpc, respectively. 
The mean alignment angles for these three bins are 
$\langle\theta\rangle = 59.07^\circ \pm 0.22$, 
$56.58^\circ \pm 0.37$, 
and $57.47^\circ \pm 0.19$, 
with corresponding alignment indices of 
$\mathcal{I} = 0.74$, $0.92$, and $0.83$, respectively. 
Groups located closer to filaments ($d_\mathrm{fila} < 1$~Mpc) therefore exhibit the largest mean angle and the smallest alignment index, 
indicating a stronger excess of perpendicular alignments compared to the random distribution. 
As the distance increases, the perpendicular alignment weakens, consistent with enhanced filamentary influence in the immediate vicinity of filaments.

The right panel explores the redshift evolution of the alignment signal, displaying results for 
$z = 0, 1, 2,$ and $3$. 
The corresponding mean alignment angles are 
$\langle\theta\rangle = 59.07^\circ \pm 0.22$, 
$59.00^\circ \pm 0.23$, 
$59.89^\circ \pm 0.30$, 
and $59.85^\circ \pm 0.49$, 
with alignment indices of 
$\mathcal{I} = 0.74$, $0.71$, $0.68$, and $0.66$, respectively. 
These results demonstrate that the perpendicular alignment shows no significant dependence on redshift: 
the signal persists across all epochs, with both $\langle\theta\rangle$ and $\mathcal{I}$ remaining nearly constant. 
This indicates that the coupling between group spin and the cosmic web is established early and remains stable throughout cosmic time.

\begin{figure}
% \plotone{f5.pdf}
\includegraphics[width=0.96\linewidth]{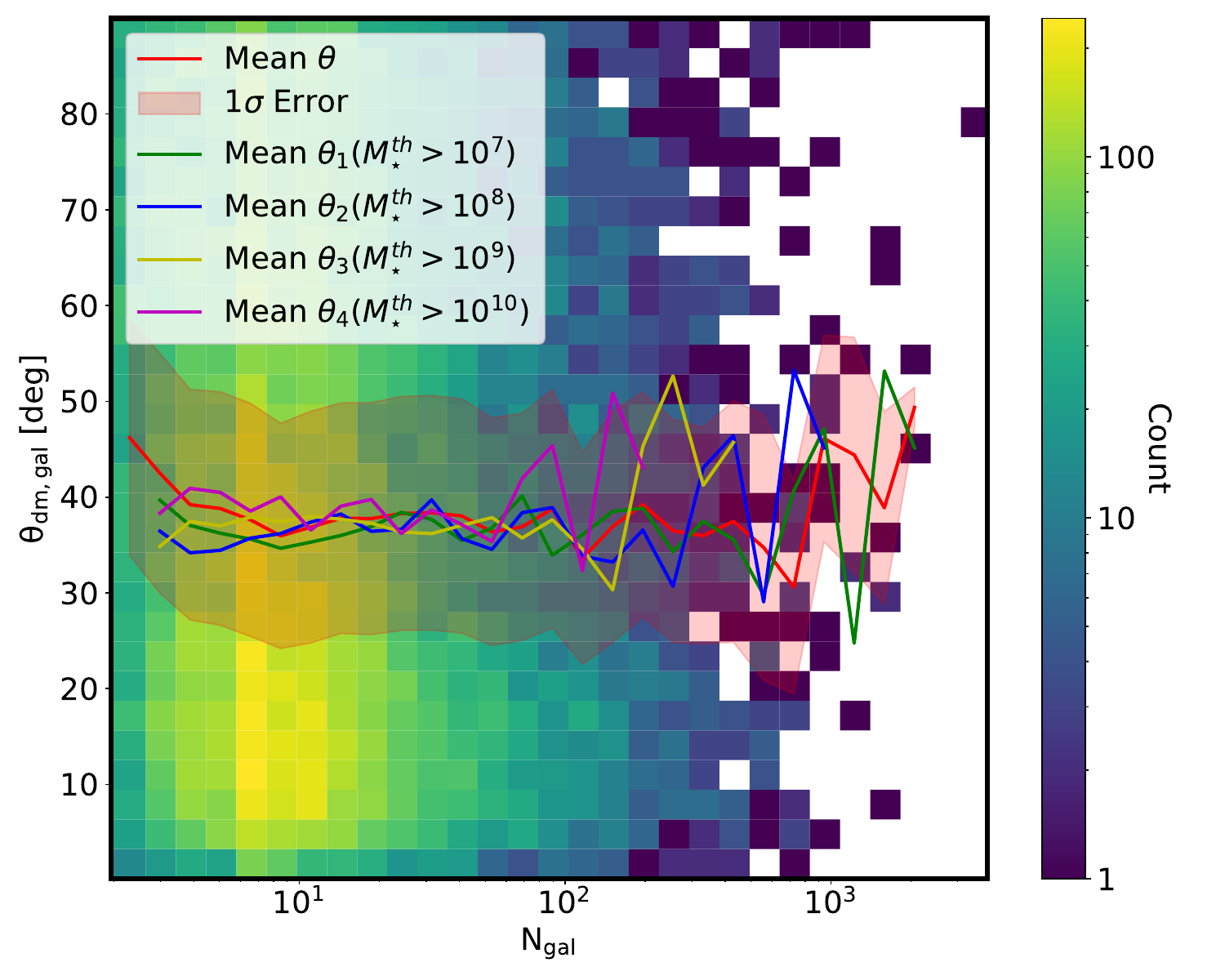}
\caption{The mean angle between the group spin direction calculated using all dark matter particles and that estimated from member galaxies, as a function of the number of member galaxies ($N_\mathrm{gal}$). Different colored lines represent results obtained by applying various stellar mass cuts to the member galaxies. Error bars indicate the standard error of the mean in each bin. The color scale indicates the number of groups (counts) in each bin. The comparison demonstrates that the typical misalignment between the two spin measurements is not strongly dependent on group richness or the specific mass cut applied to the member galaxies.}
\label{fig:f5}
\end{figure}

\subsection{Reliability of Member Galaxies in Tracing Group Spin}
As mentioned above, we have demonstrated that the spin of galaxy groups exhibits a strong perpendicular alignment with respect to nearby filamentary structures, and that the strength of this alignment depends on group mass and proximity to filaments, while remaining robust against the choice of richness and stellar mass threshold for member galaxies (see Figures~\ref{fig:f2}--\ref{fig:f4}). In terms of redshift evolution, the alignment signal shows little variation, remaining stable across cosmic time.

It is important to note that, in cosmological simulations, a galaxy group is essentially a dark matter halo, and its intrinsic angular momentum can be robustly measured using all dark matter particles within the halo. However, both in observations \citep{2025ApJ...983L...3R} and in our analysis, the group spin is typically estimated from the positions and velocities of member galaxies, which may introduce systematic differences compared to the true halo spin.

To quantify the potential bias introduced by using member galaxies as tracers, we directly compare the spin direction calculated from all dark matter particles with that estimated from the member galaxies for each group. Figure~\ref{fig:f5} presents the mean angle between these two spin measurements as a function of the number of member galaxies ($N_\mathrm{gal}$), with different colored lines corresponding to various stellar mass cuts applied to the member galaxies. The color bar indicates the number of groups in each bin. This comparison allows us to assess how well the observable group spin traces the intrinsic halo spin, and to evaluate the impact of both group richness and the adopted mass cut on the accuracy of spin estimation. The results show that, within the error bars, the typical misalignment between the two spin measurements does not exhibit a significant dependence on group richness or on the specific mass cut applied to the member galaxies. Even as the number of member galaxies increases or different mass thresholds are adopted, the angle between the spin directions estimated from dark matter particles and from member galaxies remains nearly constant, with a mean value of approximately $38^\circ$.

%%%%%%%%%%%%%%%%%%%%%%%%%%%%%%%%%%%%%%%%%%%%%%%%%%%%%
%%%%%%%%%%%%% 
%   con & dis
%%%%%%%%%%%%% 
\section{Summary}\label{sec:sum}
In this work, we have used the TNG300-1 cosmological hydrodynamical simulation \citep{2018MNRAS.480.5113M,2018MNRAS.477.1206N,2018MNRAS.475..624N,2018MNRAS.475..648P,2018MNRAS.475..676S} to systematically investigate the alignment between the spin vectors of galaxy groups and the axes of their nearest cosmic filaments. Our analysis focuses on signals extracted from numerical simulations, offering a comparison with observational data. The main results and implications of our study are summarized as follows:

\begin{itemize}
\item We confirm a strong perpendicular alignment between the spin vectors of galaxy groups and the directions of their host filaments. The mean alignment angle $\langle\theta\rangle$ is significantly larger than the random expectation, and the alignment index $\mathcal{I}$ indicates a substantial excess of perpendicular alignments. This result is consistent with previous simulation and observational studies, and supports the scenario in which group spin is primarily acquired through the orbital angular momentum of infalling galaxies along filaments.

\item The strength of the perpendicular alignment signal is found to depend systematically on group mass, and distance to filaments. More massive groups, as well as those located closer to filaments, exhibit stronger perpendicular alignment. In contrast to the dependence on mass and proximity to filaments, the alignment signal shows little dependence on redshift—remaining stable across cosmic time—and exhibits no significant trend with richness.

\item We further show that the alignment signal is robust against the choice of stellar mass threshold for member galaxies when estimating group spin. Varying the stellar mass cut does not significantly affect the measured alignment angle or alignment index.

\item We have explicitly quantified the potential bias introduced by using member galaxies as tracers of group spin, as is necessary in observational studies. By comparing the spin direction calculated from all dark matter particles with that estimated from member galaxies, we find that the typical misalignment between the two measurements is approximately $38^\circ$, and, within the error bars, does not exhibit a significant dependence on group richness or on the specific stellar mass cut applied to the member galaxies. This result demonstrates that, while member galaxies provide a practical and accessible proxy for group spin, there is an intrinsic and non-negligible offset relative to the true halo spin direction.
\end{itemize}

These results provide a robust theoretical foundation for interpreting observed spin–filament alignments and highlight the importance of considering both environmental effects and tracer selection when connecting simulations with observations.

%%%%%%%%%%%%% 
%   con & dis
%%%%%%%%%%%%% 
\section{Discussion}\label{sec:dis}
Our analysis demonstrates that the perpendicular alignment between group spin and filament direction is a robust feature in the TNG300-1 simulation \citep{2018MNRAS.480.5113M,2018MNRAS.477.1206N,2018MNRAS.475..624N,2018MNRAS.475..648P,2018MNRAS.475..676S}. Importantly, among all the factors we examined, only two -- group mass and the distance to the nearest filament -- have a significant impact on the strength of the alignment signal. Specifically, more massive groups and those located closer to filaments exhibit a stronger perpendicular alignment. In contrast, other properties such as group richness, the stellar mass threshold for member galaxies, and redshift show little or no effect on the alignment signal. This finding provides a clear and focused theoretical benchmark for interpreting observational results.

While the general trend of spin-filament alignment has been established in both simulations and observations \citep{2013ApJ...775L..42T,2012MNRAS.427.3320C,2014MNRAS.444.1453D,2021NatAs...5..742K,2021NatAs...5..839W}, our study extends this understanding by systematically isolating the key factors that influence the signal. By varying the stellar mass threshold for member galaxies and examining group richness, we show that the alignment is remarkably stable against these choices, which is reassuring for observational studies where sample completeness and richness may vary. The reason of no dependence of the group spin-filament alignment on stellar mass is leading from two points. One is the definition of group spin, which enable massive galaxies to be assigned higher mass weight even exceeding several magnitudes. Secondly, we keep the same parameter including stellar mass threshold to trace filament structure to ensure the consistency of results because of several works showing the dependence between filamentary structure and stellar mass threshold(e.g. \citep{2023MNRAS.525.4079Z}). Our direct quantification of the misalignment between the group spin measured from all dark matter particles and that inferred from member galaxies (a typical offset of about $38^\circ$) provides a practical reference for interpreting observational results, where only galaxy-based spin estimates are available.

Compared to previous simulation studies \citep{2012MNRAS.427.3320C,2019MNRAS.487.1607G}, which often focus on halo spin or use idealized tracers, our approach closely mimics observational methodologies and clarifies that, in practice, only group mass and group-filament distance need to be carefully considered when interpreting the strength of the alignment signal. This result complements recent observational findings and provides a direct simulation-based explanation for the observed trends. The importance of group mass and environment aligns with findings from \citep{2014MNRAS.444.1453D,2021NatAs...5..742K}, who also emphasized the role of mass and environment in shaping spin alignments.

A key aspect of our work is the explicit quantification of the bias introduced by using member galaxies as tracers of group spin. The nearly constant misalignment angle of $\sim38^\circ$, independent of group richness or stellar mass cut, highlights an intrinsic limitation in observational studies. While the overall trend of perpendicular alignment is preserved, this systematic offset should be considered when comparing observations with theoretical predictions. Our results also show that the alignment signal is robust against the choice of stellar mass threshold for member galaxies, suggesting that observational incompleteness or selection effects in galaxy samples are unlikely to significantly bias the inferred group spin-filament alignment. This is particularly relevant given the challenges in accurately measuring galaxy spins in observations \citep{2015ApJ...798...17Z,2015MNRAS.450.2727T}.

There are, however, several limitations to our current analysis. The use of a single simulation restricts our ability to assess the impact of different cosmological parameters or baryonic physics. Our study does not explore the detailed physical mechanisms that drive the spin-filament alignment, such as the role of mergers, anisotropic accretion, or the interplay between dark matter and baryons. Additionally, uncertainties in group identification and membership assignment, both in simulations and in real data, may affect the comparison between simulated and observed group spin.
Future studies could benefit from exploring these mechanisms using higher-resolution simulations and incorporating more sophisticated models of galaxy formation \citep{2015MNRAS.446.2744L}. In addition, the impact of different filament algorithms(e.g. \citep{2018MNRAS.473.1195L,2025ApJ...990..119G}) on the alignment of group spin-filament is also an important aspect to consider.

The finding that only group mass and group-filament distance significantly affect the alignment signal provides a practical guideline for both theoretical and observational studies. Future work could address the limitations noted above by analyzing a broader range of simulations, incorporating more sophisticated models of galaxy formation, and developing improved methods for measuring group spin in both simulations and observations. Extending the analysis to include the angular momentum of the gas and stellar components, or to study the time evolution of individual groups, would provide further insight into the origin and persistence of the alignment signal. 

Overall, our results clarify the main factors that influence the spin-filament alignment of galaxy groups and provide a useful reference for interpreting and comparing both simulation and observational studies of angular momentum in the large-scale structure.

%%%%%%%%%%%%%%%
%     Acknowledgments
%%%%%%%%%%%%%%%
\begin{acknowledgments}

%The authors thank anonymous referees for comments that substantially improved the manuscript.  
WW, PW, WHD and XXT acknowledge the financial support from the NSFC (No.12473009), and also sponsored by Shanghai Rising-Star Program (No.24QA2711100). 
This work is supported by the China Manned Space Program with grant no. CMS-CSST-2025-A03.
Y.R. acknowledges supports from the CAS Pioneer Hundred Talents Program (Category B), the NSFC grant (No. 12273037), and the USTC Research Funds of the Double First-Class Initiative. 

\end{acknowledgments}

%%%%%%%%%%%%%%%%%%%%%%
%    Bibliography
%%%%%%%%%%%%%%%%%%%%%%
\bibliographystyle{JHEP}

% \bibliography{ref.bib}

\providecommand{\href}[2]{#2}\begingroup\raggedright\endgroup

% \bibliographystyle{aasjournal}

% \acknowledgments

% \end{CJK*}
\end{document}